\documentclass[useAMS,usenatbib]{mn2e}
\usepackage{graphicx}
\usepackage{amssymb}
\usepackage{amsmath}
\usepackage{natbib}
\usepackage[section]{placeins}
\usepackage{verbatim}

\usepackage{gensymb}

\usepackage{url}

\title[A Search for Methane in the Atmosphere of GJ 1214b via GTC Narrow-Band Transmission Spectrophotometry]{A Search for Methane in the Atmosphere of GJ 1214b via GTC Narrow-Band Transmission Spectrophotometry\thanks{Based on observations made with the Gran Telescopio Canarias (GTC), installed at the Spanish Observatorio del Roque de los Muchachos of the Instituto de Astrof\'isica de Canarias, in the island of La Palma, and part of the large ESO program 182.C-2018 and the programs GTC2-10AFLO and GTC4-11AFLO.}}

\author[P. A.~Wilson et al.]
{\parbox{\textwidth}{P. A. Wilson$^{1}$\thanks{E-mail: \texttt{paw@astro.ex.ac.uk}},
K. D. Col\'on$^{2,3}$,
D. K. Sing$^{1}$,
G. E. Ballester$^{4}$,
J.-M. D{\'e}sert$^{5}$,
D.~Ehrenreich$^{6}$,
E. B. Ford$^{3,10,11}$,
J. J. Fortney$^{7}$,
A. Lecavelier des Etangs$^{6}$,
M. L{\'o}pez-Morales$^{8}$,
C. V. Morley$^{7}$,
A. R. Pettitt$^{1}$,
F. Pont$^{1}$,
A. Vidal-Madjar$^{9}$}\vspace{0.4cm}\\
\parbox{\textwidth}{$^{1}$Astrophysics Group, School of Physics, University of Exeter, Stocker Road, Exeter EX4 4QL, UK\\
$^{2}$Department of Geology and Geophysics, University of Hawai'i at Manoa, Honolulu, HI 96822, USA\\
$^{3}$Department of Astronomy, University of Florida, Gainesville, FL 32611, USA\\
$^{4}$Lunar and Planetary Laboratory, University of Arizona, Sonett Space Sciences Building, Tucson, Arizona 85721-0063, USA\\
$^{5}$Caltech Division of Geological and Planetary Sciences, Pasadena, CA 91125, USA\\
$^{6}$Observatoire astronomique de l'Universit{\'e} de Gen{\`e}ve, 51 chemin des Maillettes 1290 Sauverny, Switzerland\\
$^{7}$Department of Astronomy and Astrophysics, University of California, Santa Cruz, CA 95064, USA\\
$^{8}$Harvard-Smithsonian Center for Astrophysics, 60 Garden Street, Cambridge, MA 02138, USA\\
$^{9}$Institut d'astrophysique de Paris, CNRS; Universit{\'e} Pierre et Marie Curie, 98 bis boulevard Arago, F-75014 Paris, France\\
$^{10}$Center for Exoplanets and Habitable Worlds, 525 Davey Laboratory, Pennsylvania State University, University Park, PA 16802, USA\\
$^{11}$Department of Astronomy and Astrophysics, Pennsylvania State University, 525 Davey Laboratory, University Park, PA 16802, USA}}

\begin{document}

\date{Accepted 2013 December 3. Received 2013 December 2; in original form 2013 May 31}

\pagerange{\pageref{firstpage}--\pageref{lastpage}} \pubyear{2014}

\maketitle

\label{firstpage}

\begin{abstract}
We present narrow-band photometric measurements of the exoplanet GJ 1214b using the 10.4~m Gran Telescopio Canarias (GTC) and the OSIRIS instrument. Using tuneable filters we observed a total of five transits, three of which were observed at two wavelengths nearly simultaneously, producing a total of eight individual light curves, six of these probed the possible existence of a methane absorption feature in the 8770~--~8850~\AA\ region at high resolution. We detect no increase in the planet-to-star radius ratio across the methane feature with a change in radius ratio of $\Delta \overline{R} = -0.0007 \pm 0.0017$ corresponding to a scale height (H) change of $-0.5 \pm 1.2~H$ across the methane feature, assuming a hydrogen dominated atmosphere. We find a variety of water and cloudy atmospheric models fit the data well, but find that cloud-free models provide poor fits. These observations support a flat transmission spectrum resulting from the presence of a high-altitude haze or a water-rich atmosphere, in agreement with previous studies. In this study the observations are predominantly limited by the photometric quality and the limited number of data points (resulting from a long observing cadence), which make the determination of the systematic noise challenging. With tuneable filters capable of high resolution measurements (R~$\approx$~600--750) of narrow absorption features, the interpretation of our results are also limited by the absence of high resolution methane models below 1~$\mu$m. 
\end{abstract}


\section{Introduction}

The discovery of close-in ``super-Earths'', with masses between 1.5 and 10 M$_\oplus$, has opened an entirely new field of exoplanet research. While transiting super Earths allow the radius and mass to be measured, the regime is prone to large degeneracies between their internal and atmospheric compositions and their masses \citep{rogers10}. Characterising the atmospheres may be the only way to help constrain the overall bulk composition of these hot planets. A large planet-to-star contrast is essential when measuring transmission or emission spectra, making transiting super Earths orbiting M dwarf stars ideal for such studies.

The most studied super Earth is GJ 1214b, discovered in the MEarth ground-based transit survey \citep{charbonneau09}. GJ 1214b is a 2.7 $R_{\oplus}$ planet orbiting a M4.5 dwarf star, and therefore has a large planet-to-star radius ratio despite the stellar radius only being $0.21~R_{\sun}$. The result is a transit depth of nearly 1.5\%. The observed mass and radius of GJ 1214b are consistent with theoretical models indicative of a significant atmosphere \citep{kempton10}. Due to degeneracies in the models it is predicted that GJ 1214b is either composed of a rocky/ice core surrounded by a hydrogen-rich atmosphere, a water/ice core with an atmosphere dominated by water vapor, or a rocky core with a thin atmosphere formed by outgassing \citep{rogers10}.

Recent studies have attempted to constrain GJ 1214b's atmosphere through transmission spectroscopy \citep{charbonneau09,colon13,bean10,bean11,sada10,carter11,croll11,cross11,desert11,kundurthy11,berta12,demooij12,murgas12,narita12,narita13,fraine13,teske13}. In a majority of these studies, it has been found that GJ 1214b has a flat, featureless spectrum, with no evidence of any significant features either at optical ($\sim$600--1000 nm) or near-infrared (1.1--1.7 $\mu$m) wavelengths. It is believed that the lack of significant features supports the presence of either a heavy, metal-rich atmosphere or optically thick clouds/hazes that produce a constant level of absorption across a large range of wavelengths. One exception is a study conducted by \citet{croll11}. Specifically, they reported a significantly deeper transit depth at $\sim$2.15 $\mu$m, a wavelength where methane would be a viable source of opacity. 

To help reconcile these studies, we present narrow-band photometry of five transits of GJ 1214b, three of them around a methane absorption feature commonly found at optical wavelengths in the atmospheres of the jovian planets and Titan \citep{karkoschka94}. The observations were acquired using the tuneable filter imaging mode on the Optical System for Imaging and low Resolution Integrated Spectroscopy (OSIRIS) instrument installed on the 10.4~m Gran Telescopio Canarias (GTC). Tuneable filters (TFs) have several advantages over low resolution spectroscopy; they provide accurate differential photometry whilst also allowing for relatively high resolution measurements (R~$\approx$~600--750), compared to low-resolution grisms. The relatively high resolution has the advantage of tuning the filters to avoid prominent telluric lines (see \citealt{hanuschik03} for a high-resolution sky emission atlas). Since no diffraction gratings are used, TFs can be much more efficient \citep{colon10,sing11b}, especially for observing atomic absorption features that typically have a narrow spectral range. Combining this technique with the 10.4 m aperture of the GTC telescope makes it possible to study the atmospheres of planets orbiting fainter stars compared to the hot Jupiters HD 209458b and HD 189733b, which have bright host stars and large atmospheric scale heights, making them two best studied cases thus far.

The methane feature that we focus on is the blue edge of the methane absorption band at 8900 \AA\ and is predicted to cause additional absorption during transit at a level of $\sim$0.1\%, assuming a hydrogen-rich atmosphere \citep{berta11}. In \S\ref{obs}, we describe our observations. In \S\ref{reduction} and \S\ref{analysis}, we describe our data reduction and analysis procedures. We present our results in \S\ref{results}, where we also discuss the implications of stellar activity, equilibrium cloud models and the possible presence of methane in the atmosphere of GJ 1214b. Finally, we conclude with a summary of our findings in \S\ref{conclusion}.

\section{Observations}
\label{obs}

Photometric observations of GJ 1214b were conducted using the GTC telescope on La Palma. For all observations, we used the tuneable filter (TF) imaging mode on OSIRIS \citep{cepa03,cepa00,cepa98} to acquire photometry within a bandpass of 12 \AA. The TF imager allows for custom bandpasses with a central wavelength between 651--934.5 nm and a FWHM of 12--20 \AA\ to be specified.

Out of the five transits observed, three transits were observed in the 8770 -- 8850~\AA region by alternating between two narrow bandpasses, each with a full width at half maximum (FWHM) of 12 \AA, allowing us to perform near simultaneous photometry at two wavelengths. Observing one transit at two wavelengths simultaneously allows for a more accurate comparison between two wavelengths as systematic variations caused by varying weather conditions or stellar activity are likely to affect both light curves similarly. For the observations done at two wavelengths we specifically chose our bandpasses so that one was located in the continuum, at a shorter wavelength of 8770 \AA\ and 8784.5 \AA\ compared to the other band located at 8835 \AA\ and 8849.6 \AA, within the methane absorption band. As described in \cite{colon10} and \cite{sing11b}, another property of the TF imaging mode is that the effective wavelength decreases radially outward from the optical center, so we attempted to position the target and a single ``primary'' reference star (i.e., one most comparable in brightness to the target) at the same distance from the optical center so as to observe both stars at the same wavelengths.\footnote{Due to technical issues, the positioning for some of the observations was not as expected, and the target and a single reference star were not always observed at the same exact wavelengths. See \S \ref{sky_lines}.} The other reference stars were thus observed at slightly different wavelengths than the target, due to their different distances from the optical center.

All observations were performed with $1\times1$ binning and a fast pixel readout rate of 500 kHz, a gain of 1.46~$e^{-}$/ADU and a read noise of $\sim8~e^{-}$ as well as a single window located on one CCD chip. The size of the window varied for each observation, but was chosen to be large enough so as to contain the target and several reference stars of similar brightness. Data points with analog-to-digital unit (ADU) counts larger than 45,000 were removed to ensure the measurements were taken in the linear regime of the CCD detector. The data presented in this paper originated from two separate observing programs by PI. D. Sing (ESO program 182.C-2018 see \S\ref{transit-8100} and \S\ref{transit-8550}) and PI. K. Col\'on (GTC2-10AFLO and GTC4-11AFLO see \S\ref{transit-8770-I}, \S\ref{transit-8770-II} and \S\ref{transit-8785}) and each had slightly different observing strategies. Further details regarding each specific transit observation are given in the following sections.

\subsection{8100 \AA\ Transit, 17 August 2010}
\label{transit-8100}

Observations of the 2010 August 17 transit were tuned to a target wavelength of 8100 \AA, with the target 3.7 arc minutes away from the optical centre. The observations began at 21:18 UT and ended at 23:29 UT, during which time the airmass ranged from 1.11 to 1.44. Due to variable seeing, ranging from 0.7 to 1.2\arcsec, the telescope was defocused to avoid saturation. Two reference stars were selected (more on the selection technique in \S\ref{reduction}). The observations were windowed using a 1160$\times$760 pixel section on CCD1. Twelve images containing counts greater than 45 000 ADU were removed to ensure linearity. The exposure time was kept at 60~s throughout the sequence, with a corresponding $\sim$12~s of readout time.

\subsection{8550 \AA\ Transit, 2 June 2010}
\label{transit-8550}

Observations of the 2010 June 2 transit were tuned to a target wavelength of 8550 \AA, with the target 1.3 arc minutes from the optical centre. The observations began at 23:48 UT and ended at 03:04 UT, during which time the airmass ranged from 1.27 to 1.87. Due to a technical problem with the secondary mirror, re-focusing was not possible during the whole sequence. This caused an increase in the full width at half-maximum (FWHM) of the Point Spread Function (PSF) of the stars from 0.9 to 1.9~\arcsec, resulting in a notable decrease in the peak counts levels. Three reference stars were selected. The observations were performed using CCD1 and no windowing was done. One data point with counts greater than 45 000 ADU was removed to ensure linearity (see \S~\ref{reduction}). The exposure time was kept at 120~s throughout the sequence, with a corresponding $\sim$24.5~s of readout time for the full frame.

\subsection{8770 and 8835 \AA\ Transit, 28 August 2010}
\label{transit-8770-I}

Observations of the 2010 August 28 transit were tuned to the target wavelengths of 8770 and 8835 \AA, with the target 3.2 arc minutes from the optical centre. The observations were done by alternating between the two wavelengths in sets of two exposures at each wavelength. The seeing was variable throughout the observations, so the telescope was defocused and the exposure time was changed to avoid saturation. The observations were done in queue (service) mode. The exposure time started at 100 s and was later increased to 150 s and then to 200 s towards the end of the observations.  The observations began at 22:00 UT and ended at 00:30 UT, during which time the airmass ranged from 1.26 to 2.32 and the FWHM varied between 1.0 to 2.6~\arcsec. There are some small gaps in the data towards the beginning of the observations due to minor technical issues. Also, there is some vignetting in the last few images due to the low elevation of the telescope, so we exclude these from our analysis. Two reference stars were selected. Two data points with counts greater than 45 000 ADU were removed to ensure linearity. The observations were windowed using a 850$\times$4102 pixel section on CCD2 with a corresponding $\sim$22~s of readout time.

\subsection{8770 and 8835 \AA\ Transit, 10 June 2011}
\label{transit-8770-II}

\begin{figure}
\includegraphics[width=84mm]{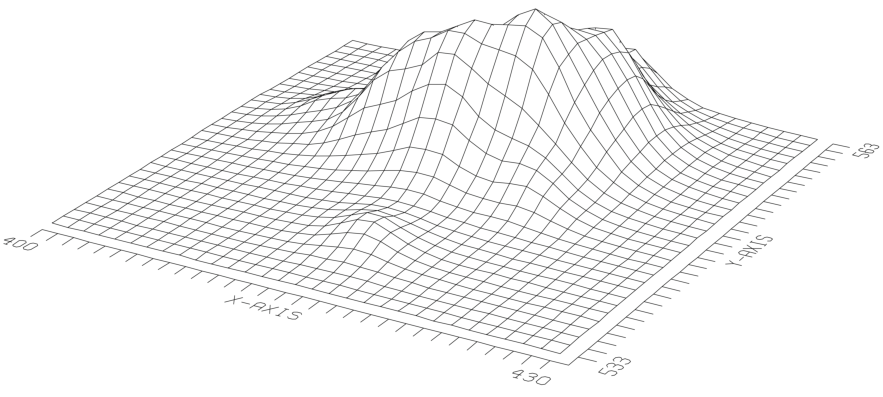}
 \caption{A surface plot of GJ 1214 showing an uneven PSF due to a misalignment of the M1 mirror during the 8770 and 8835 \AA\ observations on the 10$^{\rm{th}}$ of June 2011 (see Section \S~\ref{transit-8770-II}). The small bump seen towards the front of the larger PSF is due to one of the hexagonal mirrors not being properly aligned. By choosing a larger photometry aperture, the effects caused by the distorted PSF were removed.}
  \label{surface}
\end{figure}

Observations of the 2011 June 11 transit were tuned to the target wavelengths of 8770 and 8835 \AA, with the target 3.2 arc minutes from the optical centre. The observations were done by alternating between the two wavelengths in sets of two exposures at each wavelength. The conditions were clear, and observations took place during bright time in visitor mode. Observations began at 23:40 UT and ended at 02:48 UT, during which time the airmass ranged from 1.09 to 1.19 and the FWHM varied between 1.3 and 2.2~\arcsec. The observations started 25 min later than planned because one of the M1 mirror segments was found to be slightly misaligned (see Fig. \ref{surface}). One segment of the mirror would not stack with the other segments. Attempts were made to correct this, although the problem persisted throughout the observations. As this problem had the same effect on all the stars (i.e., each star had an extended PSF, see Fig. \ref{surface}), we have assumed the photometry was not significantly affected by this problem since we chose a larger aperture that included the photons from the unstacked segment. Three reference stars were selected. The observations were windowed using a 850$\times$3250 pixel section on CCD1. An exposure time of 100~s was used throughout the sequence, with a corresponding $\sim$19~s of readout time.

\subsection{8784.5 and 8849.6 \AA\ Transit, 21 July 2010}
\label{transit-8785}

Observations of the 2010 July 22 transit were tuned to the target wavelengths of 8784.5 and 8849.6 \AA, with the target 2.9 arc minutes from the optical centre. The observations were done by alternating between the two wavelengths in sets of two exposures at each wavelength. The conditions were clear and the observations took place during bright time in queue mode. The observations began at 00:26 UT and ended at 02:11 UT. The airmass ranged from 1.25 to 1.87. The actual seeing varied between 0.9 and 1.4~\arcsec. A slight defocus was used in order to avoid saturation. Two reference stars were selected. The observations were windowed using a 849$\times$3774 pixel section on CCD2. An exposure time of 120~s was used throughout the sequence, with a corresponding $\sim$22~s of readout time.

\begin{figure}
\includegraphics[width=84mm]{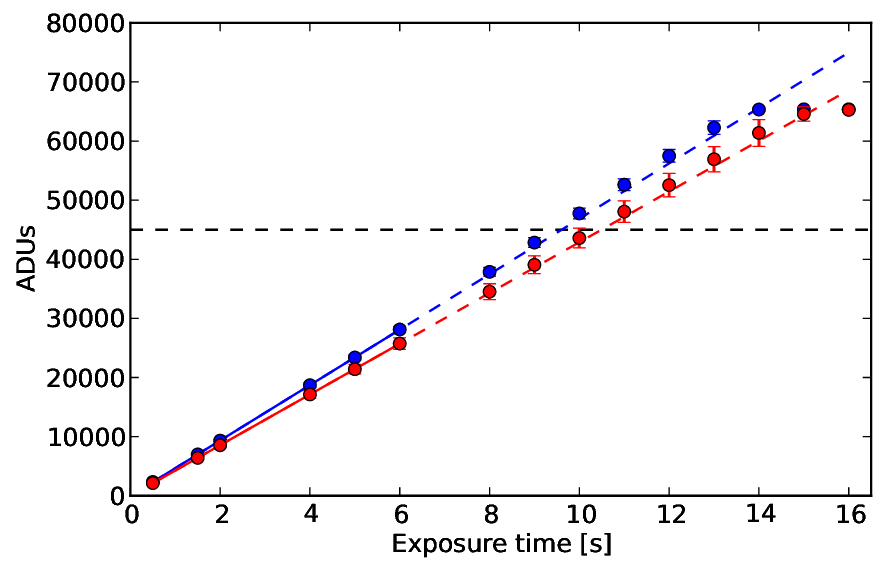}
 \caption{The OSIRIS CCD1 (red) and CCD2 (blue) 500kHz exposure curves showing the linearity of the detector. The measurements connected by a solid line were used to fit the estimated linearity of the detector. Images which had the brighter reference star at more than 45,000 ADUs were removed in order to ensure they used data sampled within the linear regime of the CCD.}
  \label{exp-curve}
\end{figure}

\section{Data Reduction}
\label{reduction}

For all our data sets, we used standard IRAF\footnote{IRAF is distributed by the National Optical Astronomy Observatories, which are operated by the Association of Universities for Research in Astronomy, Inc., under cooperative agreement with the National Science Foundation.} procedures for bias subtraction and flat-field correction. For the flat-field correction, we used dome flats that were taken after each observation and for each filter setting. For the observations done at the methane probing wavelengths $8770$~\AA\ (two transits), $8784.5$~\AA, $8835.0$~\AA\ (two transits) and $8849.6$~\AA\ were affected by the presence of sky lines. We therefore performed a sky subtraction of these images using the IRAF package TFred\footnote{Written by D. H. Jones for the Taurus Tunable Filter, previously installed on the Anglo-Australian Telescope; \url{http://www.aao.gov.au/local/www/jbh/ttf/adv_reduc.html}}.

Aperture photometry was done using the APPHOT package in IRAF. To obtain the best possible photometry, a large number of apertures and sky annuli were explored. The aperture and sky annulus combination which produced the least amount of scatter in the continuum (lowest $\chi^2$ value by fitting a straight line to the continuum) was chosen. The number of reference stars varied depending on the size of the CCD readout window, the location of the sky lines as well as the observed scatter in the photometry of each reference star. To determine the optimal number of reference stars all stars above 15,000 ADU were initially chosen as potential reference stars. Each star which did not help reduce the overall scatter in the continuum, such as fainter stars affected by the sky emission rings (see \S\ref{sky_lines}) were removed. 

The linearities of the CCD1 and CCD2 detectors were evaluated by measuring the average ADU counts of centrally windowed flat field images as a function of exposure time (see Fig. \ref{exp-curve}). Using the measured points known to be within the linear regime of the CCD ($< $~25,000~ADU), a linear extrapolation of ADUs as a function of exposure time was created. To ensure the observations were not affected by non-linearity effects, the few images that contained a reference star with more than 45,000~ADUs were discarded, as counts above this level were shown to deviate from the linear extrapolation by more than 1~$\sigma$ on CCD2. The resulting light curves are shown in Fig. \ref{transits-sing} and \ref{transits-colon}.

\begin{figure}
\includegraphics[width=84mm]{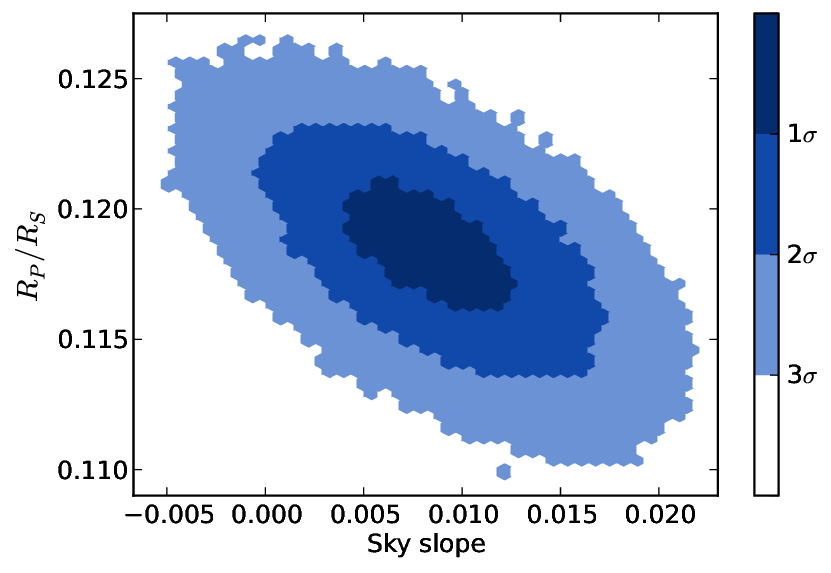}
 \caption{The posterior MCMC distribution showing the relationship between the radius ratio and the slope of the sky term for the 8770.0~\AA\ observations on the 10$^{\rm{th}}$ of June 2011. The different shadings represents the 1$\sigma$  (dark-blue), 2$\sigma$  (mid-blue) and 3$\sigma$ (light-blue) confidence intervals.}
  \label{mcmc}
\end{figure}

\section{Analysis}
\label{analysis}
\subsection{Light curve fits}

\begin{figure*}
\includegraphics[width=154mm]{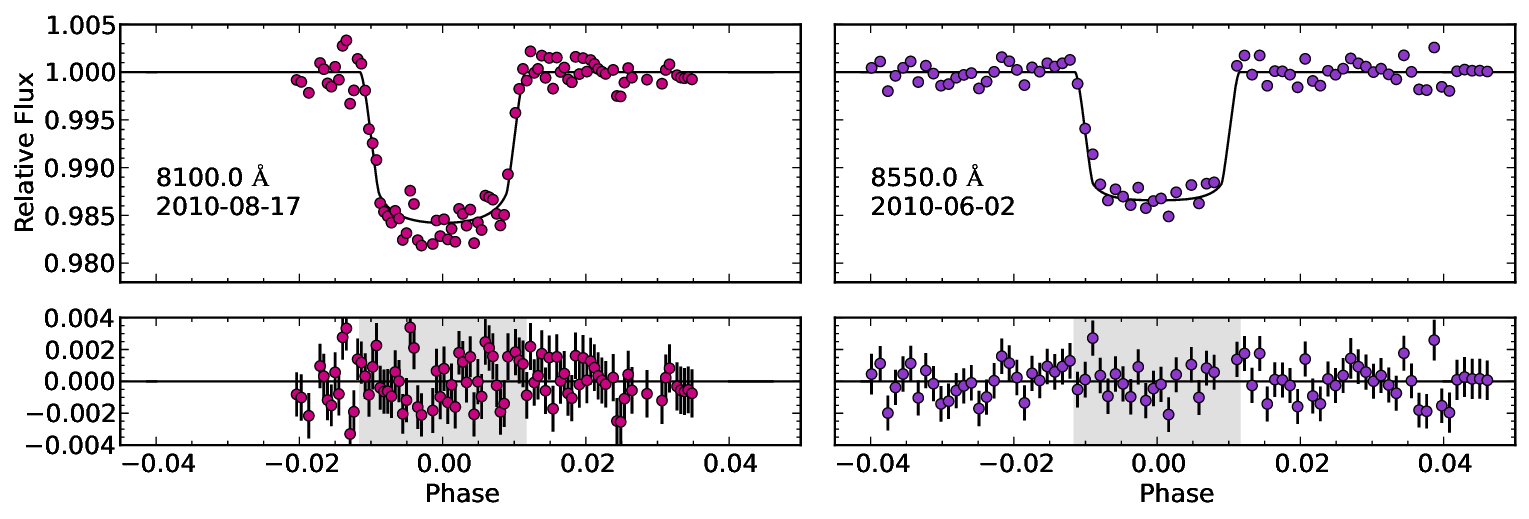}
 \caption{The GTC OSIRIS narrow band light curves with the target wavelength tuned to $8100.0$ \AA\ (left) and $8550.0$ \AA\ (right). Below each light curve are the residuals from the best fit. The $8550.0$ \AA\ light curve has a considerable shallower transit depth compared to the other transits. This could be due, in part, to a below average number of star spots on the surface of GJ 1214 effectively creating a shallower transit depth.}
  \label{transits-sing}
\end{figure*}

\begin{figure*}
\centering
\includegraphics[width=154mm]{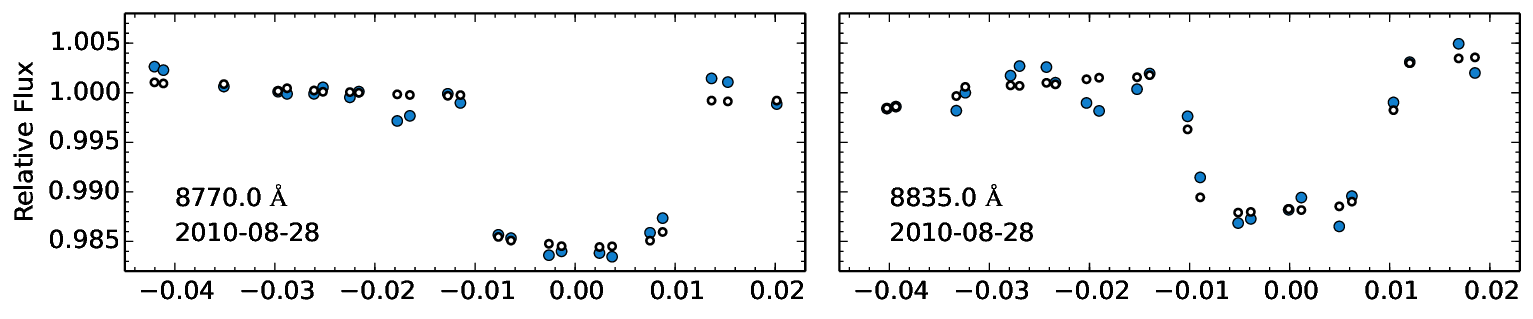}
\includegraphics[width=154mm]{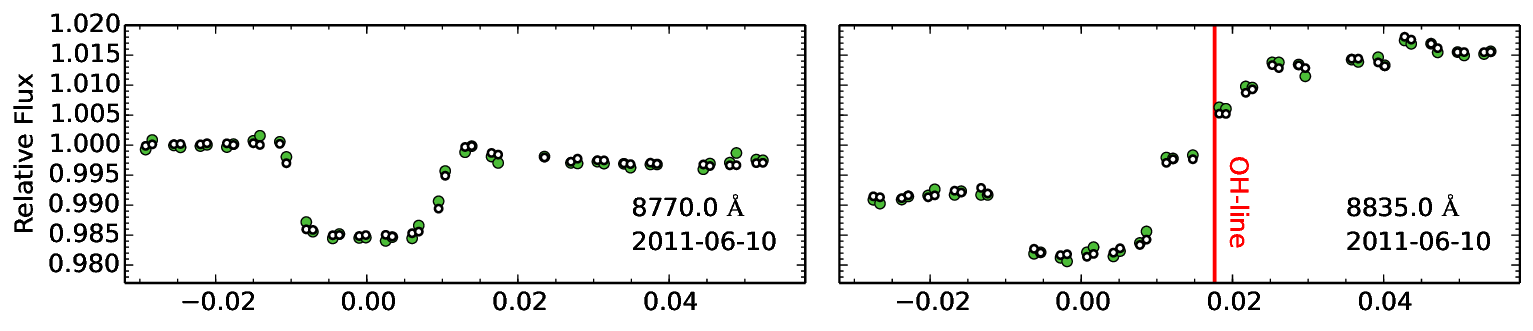}
\includegraphics[width=154mm]{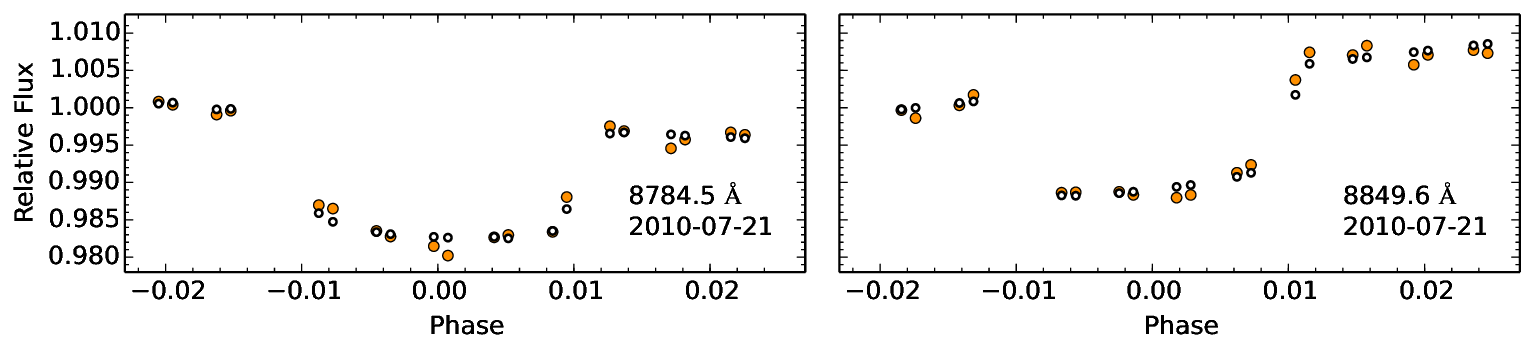}
 \caption{Shown are the raw (non-detrended) transit light curves at the off-methane target wavelengths $8770$~\AA\ (two transits) and $8784.5$~\AA\ (on the left) and the methane probing target wavelengths $8835.0$~\AA\ (two transits) and $8849.6$~\AA\ (on the right). The hollow white points represent the best-fit model. The red vertical line shows the phase during which the $8835$~\AA\ 10$^{\rm{th}}$ of June 2011 transit shows a wavelength drift across a strong OH emission line doublet near 8829.5\AA\ (see \S~\ref{wave_shift}).}
  \label{transits-colon-raw}
\end{figure*}

\begin{figure*}
\centering
\includegraphics[width=154mm]{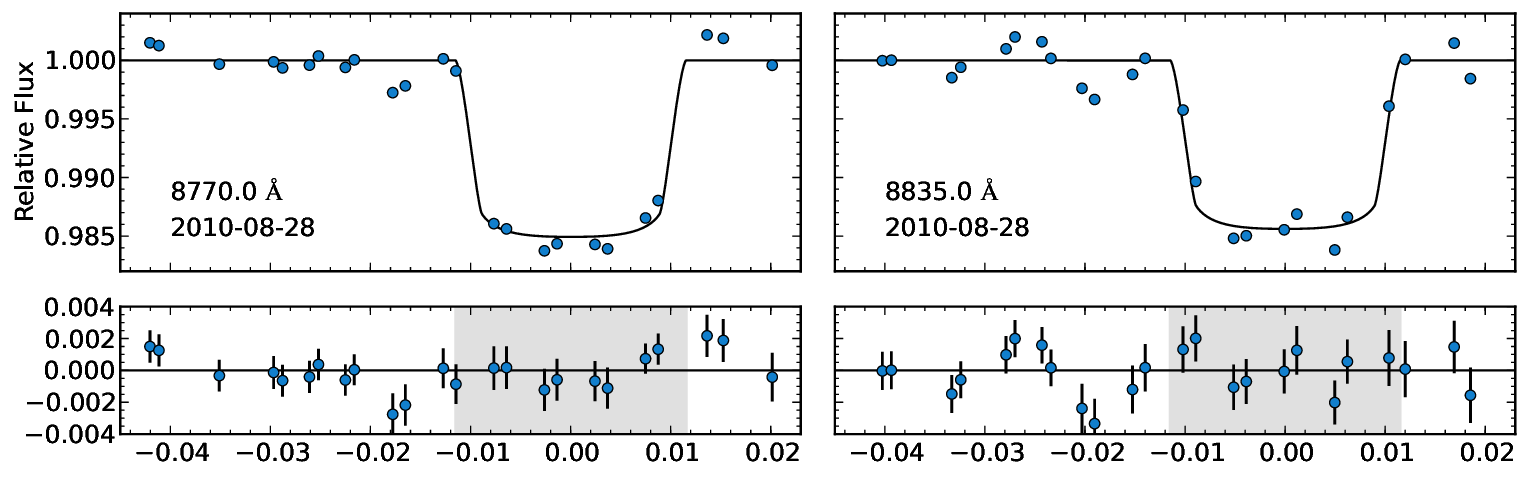}
\includegraphics[width=154mm]{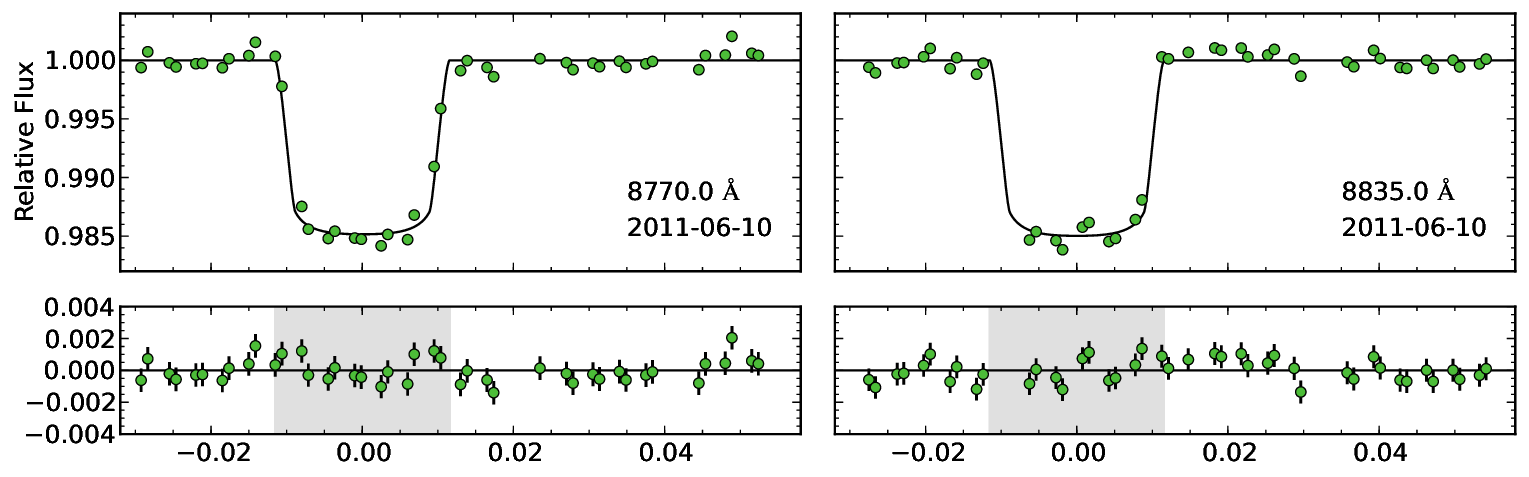}
\includegraphics[width=154mm]{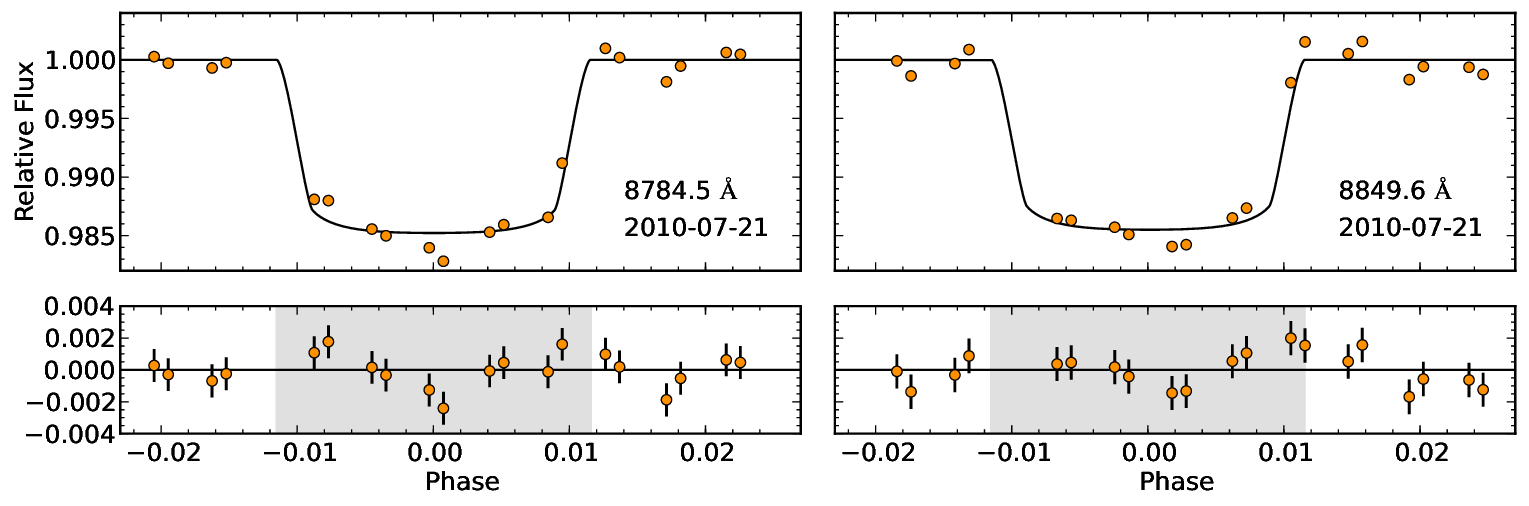}
 \caption{The GTC OSIRIS narrow band light curves at the off-methane target wavelengths $8770$~\AA\ (two transits) and $8784.5$~\AA\ (on the left) and the methane probing target wavelengths $8835.0$~\AA\ (two transits) and $8849.6$~\AA\ (on the right). Below each light curve are the residuals from the best fit.}
  \label{transits-colon}
\end{figure*}

\begin{figure}
\centering
\includegraphics[width=84mm]{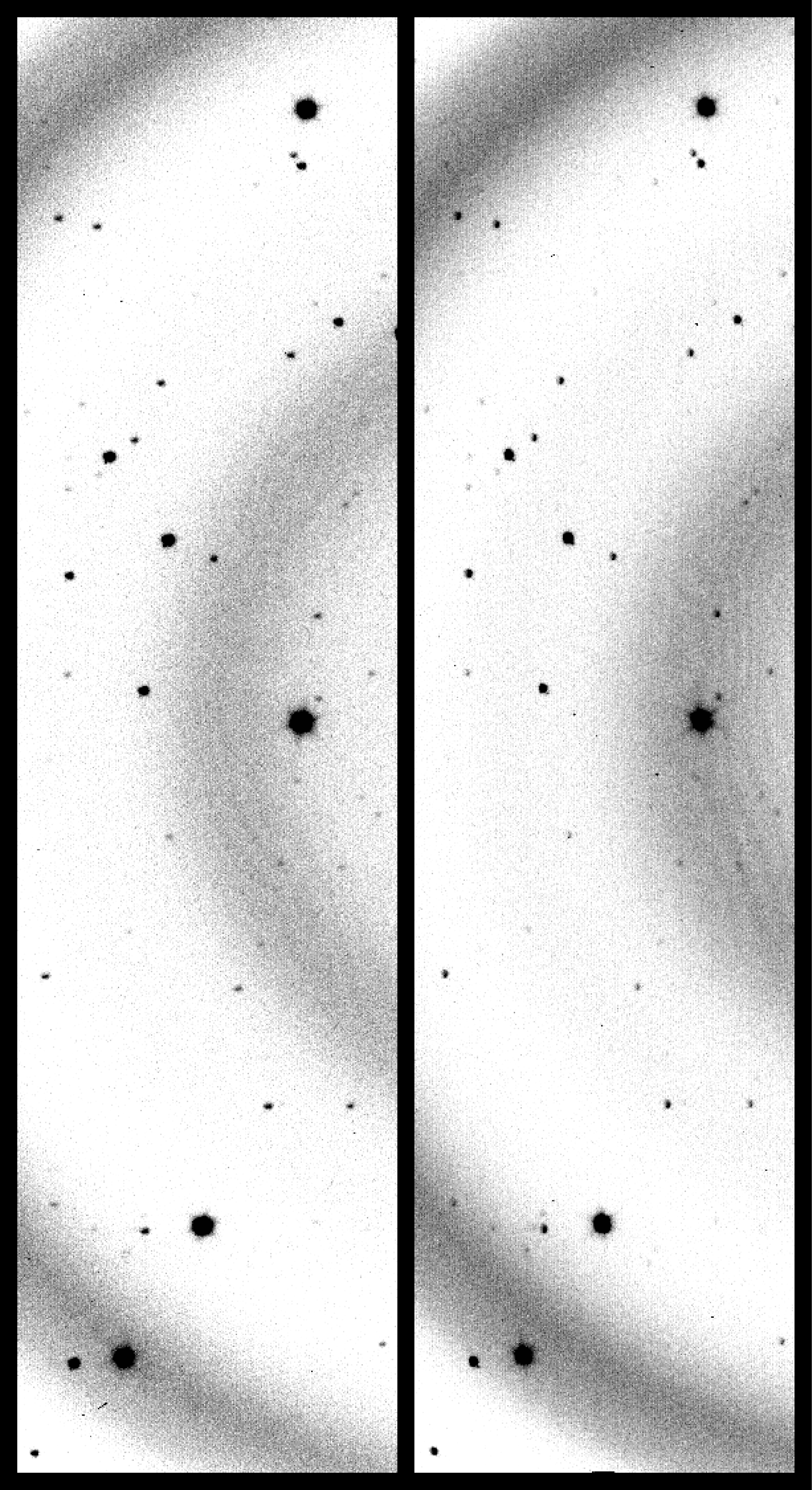}
 \caption{These two images exemplify the drift in OH sky emission observed during an observing sequence. The image on the left was taken at the beginning of the sequence, whilst the image on the right was taken towards the end of the sequence during the 8770.0~\AA\ observations on the 10$^{\rm{th}}$ of June 2011. For some of the observations a linear correlation with sky line position was found and used to detrend the systematic effects it was causing in the data.}
  \label{OB-ring}
\end{figure}

The transit light curves were fitted using the analytical transit equations of \cite{mandel02}. The best fit light curve, together with the uncertainties associated with the fits, were determined by performing a Markov chain Monte Carlo algorithm (MCMC); see, \cite{gregory05} for the use of MCMC in uncertainty estimates, \cite{collier07} for the application to transit fitting, and \cite{pont09} for our specific implementation. This gave us a posterior probability distribution which we used to define the uncertainties (see \textsection \ref{noise} and Fig. \ref{mcmc}). {\bf For a discussion on how the short baselines affect the radius ratio uncertainties, we refer the reader to Appendix~\ref{appendix}}. Individual light curve fits were generated for each transit corresponding to different wavelengths. The initial starting parameters were from \citet{bean11}; see below. We used 5 chains each consisting of 500,000 steps, trimming away the first 50,000 points with a $\sim$25\% of the proposed parameter steps being accepted. 

The free parameters in the fit were the radius ratio, $R_p/R_s$ and the sky-ring positions outlined in \S\ref{sky_lines} and summarised in Table. \ref{system-parameters}. The fixed parameters were the period $P=1.58040481$~days, mid-transit time $T_0=2454966.525123$~BJD$_{\rm{TDB}}$ (see Table \ref{system-parameters} for calculated ephemerides), impact parameter $b=0.27729$ and the quadratic limb-darkening coefficients, $u_1$ and $u_2$, which varied depending on the wavelength of the observations. The stellar and orbital parameters were also kept fixed with $R_s= 0.21~R_{\sun}$, the eccentricity $e=0$ and the scaled semimajor axis $a/R_s=14.9749$. We fix these values to allow for a more accurate comparison with \cite{bean10}, \cite{desert11}, \cite{croll11} and \citet{berta12}.

The limb darkening coefficients used were calculated using the ATLAS stellar atmospheric models\footnote{\url{http://kurucz.harvard.edu/grids.html}} following \cite{sing10} and are listed in Table \ref{system-parameters}. A quadratic limb darkening law of the following form was used

\begin{equation}
\frac{I(\mu)}{I(1)} = 1-u_1(1-\mu)-u_2(1-\mu)^2,
\end{equation}
where $I(1)$ is the intensity at the centre of the stellar disk, $\mu = \cos(\theta)$ is the angle between the line of sight and the emergent intensity while $u_1$ and $u_2$ are the limb darkening coefficients.

\begin{table*}
\centering
\caption{System parameters for GJ 1214b$^{\rm{a}}$.}
\label{system-parameters}
 \begin{tabular}{@{}lcccccccc}
  \hline
  \hline
  Wavelength & $R_{\mathrm{pl}}/R_s$ & BJD$_{\rm{TDB}}$ & Date &$u_1$ & $u_2$ & Trends & $\sigma_1$ & $\sigma_r$\\
  \hline
  $8100.0$ \AA 	&  $0.12038\pm0.0013$ & 2455426.422923	& 2010-08-17	& $0.1797$ 	& $0.3200$	& none	&  $0.00141$ & $0.00020$\\ 
  $8550.0$ \AA 	&  $0.11042\pm0.0014$ & 2455350.563492	& 2010-06-02	& $0.0552$ 	& $0.3029$	& none	&  $0.00107$ & $0.00028$\\
  $8770.0$ \AA 	&  $0.11843\pm0.0025$ & 2455437.487593	& 2010-08-28	& $0.0552$ 	& $0.3243$	& sky ring position		&  $0.00116$ & $0.00042$\\
  $8770.0$ \AA 	&  $0.11754\pm0.0016$ & 2455723.539027	& 2011-06-10	& $0.0552$ 	& $0.3243$	& sky ring position		&  $0.00073$ & $0.00034$\\
  $8784.5$ \AA 	&  $0.11724\pm0.0020$ & 2455399.556041	& 2010-07-21	& $0.0556$ 	& $0.3266$	& sky ring position		&  $0.00102$ & $0.00043$\\
  $8835.0$ \AA 	&  $0.11556\pm0.0032$ & 2455437.487593	& 2010-08-28	& $0.0577$	& $0.3357$	& sky ring position		&  $0.00140$ & $0.00048$\\
  $8835.0$ \AA 	&  $0.11791\pm0.0016$ & 2455723.539027	& 2011-06-10	& $0.0577$ 	& $0.3357$	& sky ring position		&  $0.00071$ & $0.00037$\\
  $8849.6$ \AA 	&  $0.11595\pm0.0024$ & 2455399.556041	& 2010-07-21	& $0.0584$	& $0.3387$	& none	&  $0.00107$ & $0.00049$\\
  \hline
 \end{tabular}
 \linebreak[4]
\linebreak[4]
{\footnotesize $^{\rm{a}}$ Ephemeris from \protect\cite{bean11} with $P = 1.58040481 \pm 1.2\rm{E}\text{-}7$~days and T$_c = 2454966.525123 \pm 0.000032$ BJD$_{\rm{TDB}}$.}\\
\end{table*}

\subsection{The effects of Earth's Atmosphere}
\label{earth}
In an attempt to assess the photometric variability caused by the Earth's atmosphere we studied the ratio of the reference star fluxes as a function of time and looked for correlations such as detector position, airmass, FWHM and the position of the OH emission sky lines present in the data (see Fig.~\ref{OB-ring}). In order to determine which correlations were significant, the Bayesian Information Criterion (BIC) was computed and the model with the lowest BIC was accepted. The position of the OH sky emission rings, which were only visible in the images observed around the methane feature, were the dominant systematic effect occurring when a sky ring drifted across either the target or the reference stars. The sky-position was found to behave linearly throughout the observing sequence and was modelled by a linear fit to the position of the sky rings. For the data with no sky lines present, a slope term was used to correct for the slope of the out-of-transit flux. The effects of airmass and FWHM were present, but not strong enough to warrant any detrending. In the case of the slight FWHM trends, they seemed to mainly influence the data under variable seeing conditions. The light curves with their associated correlations were all fit simultaneously by performing a MCMC. A summary of the results can be seen in Table \ref{system-parameters}. 

\begin{figure}
\centering
\includegraphics[width=84mm]{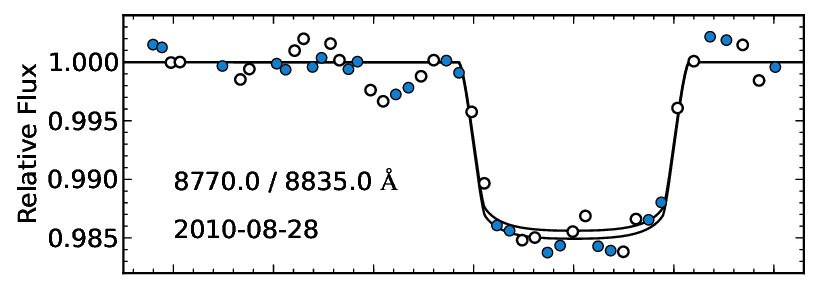}
\includegraphics[width=84mm]{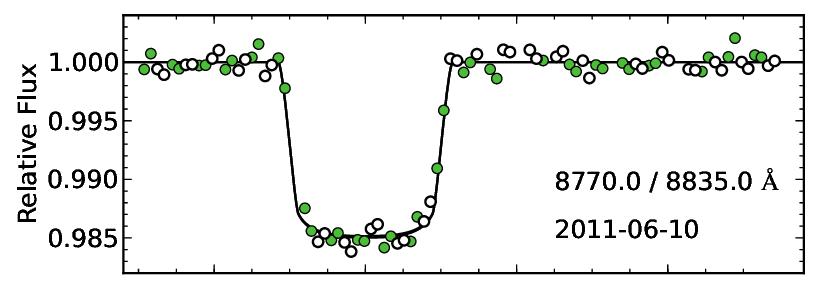}
\includegraphics[width=84mm]{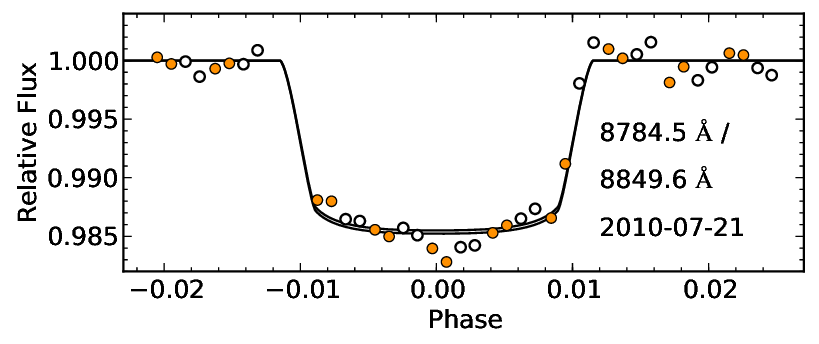}
 \caption{The GTC OSIRIS narrow band light curves with the light curves obtained on the same night over plotted with their best fit models to enable direct comparison of the transit depth differences. The white points indicate the longer wavelength within each sequence.}
  \label{transits-overplot}
\end{figure}

\subsection{The presence of sky lines}
\label{sky_lines}
When using the tuneable filters on the OSIRIS instrument the observed wavelength decreases radially outward from the center of the tuneable filter due to a difference in the optical path length that the light has to travel. This effectively causes most of the stars in the field to be observed at slightly different wavelengths. The functional form of this radial wavelength dependance can be described using the following equation:

\begin{equation}
\lambda (r) = \lambda_0 - 5.04\times r^2 + a(\lambda)\times r^3
\label{ring-eq}
\end{equation}

\noindent where the chromatic colour term is expressed as $a(\lambda)~=~6.1781-1.6024\times10^{-3}\times \lambda + 1.0215\times10^{-7}\times \lambda^2$, where $ \lambda_0$ (\AA) is the central wavelength at which the tuneable filter is tuned, and $r$ (arcmin) is the distance outward from this centre. The effects of the radial wavelength dependence is easily seen in Fig. \ref{OB-ring} where the prominent OH sky lines are visible. Ideally, each exposure is taken at the same wavelength throughout an observing sequence within a tolerance typically of 1--2~\AA, however, this is not the case for these observations. At the methane probing wavelengths $8770$ \AA, 8784.5~\AA\ and $8835.0$~\AA\ a clear drift in wavelength is observed (see Fig.~\ref{shift}). This causes systematic effects in the observations when a significant portion of the sky-ring crosses either the target star or the reference stars. A linear combination of sky-ring position of the form $A\times~sky~+~B$ was multiplied by the light curve fit, with $sky$ being the sky ring position with $A$ and $B$ being parameters set to vary freely in order to detrend this systematic effect. For the observations at 8100~\AA, 8550~\AA\ and 8849.6~\AA\ there were no interfering sky lines present in the data.

\subsection{Noise Estimate}
\label{noise}
The resulting light curves shown in Fig. \ref{transits-sing} and Fig. \ref{transits-colon} are affected by both white noise (noise uncorrelated with time, such as photon noise) as well as red noise (noise which correlates with time, such as airmass). In order to obtain realistic uncertainties the red noise must be taken into account, since only using Poisson noise can underestimate the uncertainties. We estimated the level of white noise ($\sigma_w$) and red noise ($\sigma_r$) using techniques described in \citet{pont06}. 
The relationship between $\sigma_w$ and $\sigma_r$ is given by

\begin{equation}
\sigma^2_1 = \sigma^2_w + \sigma^2_r
\end{equation}

\noindent where $\sigma_1$ is the standard deviation of the unbinned residuals, i.e., the difference between the individual normalised flux measurements and the best fit models of the transit light curves. In the absence of red noise the standard deviation in the binned residuals is expressed as

\begin{equation}
\sigma_N = \frac{\sigma_1}{\sqrt{N}}\sqrt{\frac{M}{M-1}}
\label{sig_n}
\end{equation}

\noindent where M is the number of bins each containing N points. However, since $\sigma_N$ is in most cases larger than the above calculated value (Eq. \ref{sig_n}) due to the presence of red noise, the effects of red noise on the radius ratio had to be taken into account by using a re-weighting factor. The contribution by red noise was estimated by choosing N to be equal to the number of points in the transit, which varied in accordance with the cadence of the observations and can be written as

\begin{equation}
\sigma_r = \sqrt{\sigma_N^2-\frac{\sigma_w^2}{N}}.
\end{equation}

\noindent The red noise was then used to recompute the error bars, taking systematic noise into account by using a re-weighting factor, $\beta$, expressed as

\begin{equation}
\beta = \frac{\sigma_r}{\sigma_w} \sqrt{N}.
\label{beta}
\end{equation}

\noindent The individual parameters used in the light curve fitting together with the estimated white and red noise are summarised in Table \ref{system-parameters}.

\section{Results and Discussion}
\label{results}

\begin{figure}
\includegraphics[width=84mm]{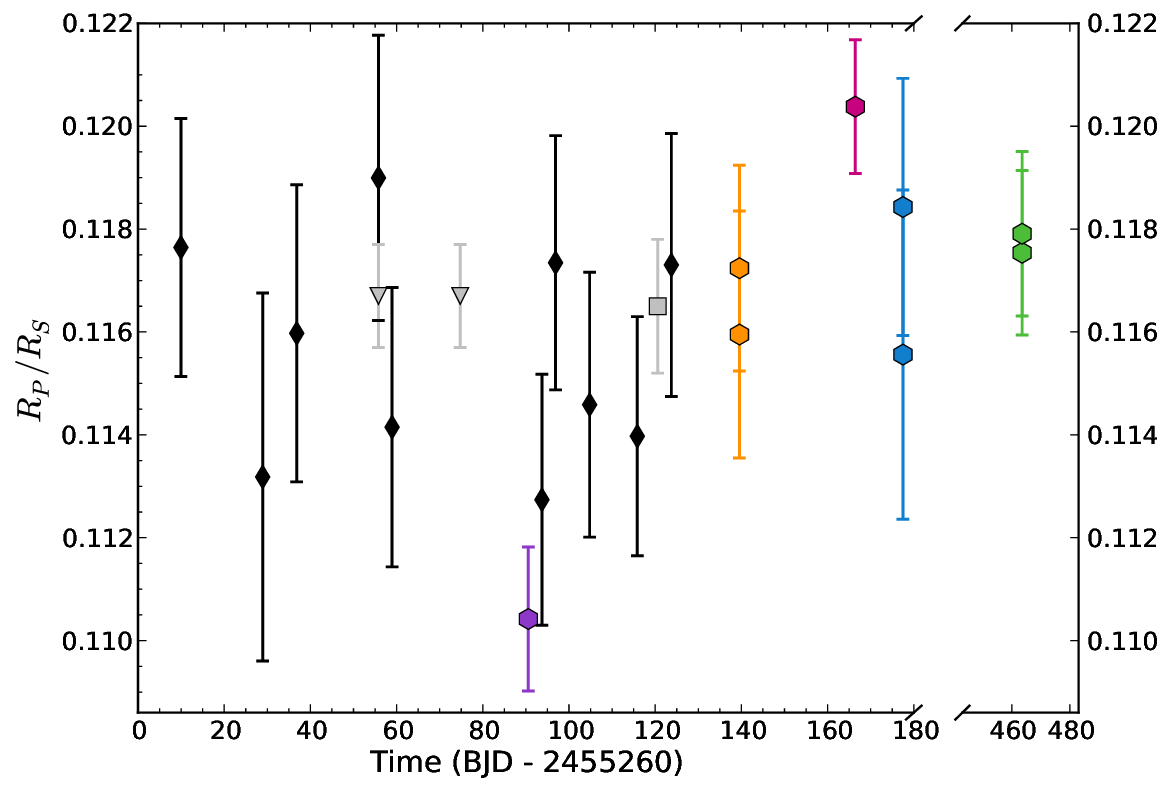}
 \caption{Shown are the radius ratios as a function of time (BJD), spanning a time period from the 29$^{\rm{th}}$ of May, 2009 until 10$^{\rm{th}}$ of June 2011. The black diamonds are measurements by \protect\cite{carter11} in the Sloan z' band using the 1.2 meter FLWO telescope. The grey triangles represent the VLT measurements by \protect\citep{berta11} in z' band. The grey squares represents the \protect\cite{demooij12} measurement with GROND instrument at the 2.2~m MPI/ESO telescope in the z band. The coloured hexagonal markers, represent our data within the z band wavelength range. The colours of the hexagonal points correspond to those in Fig.~\ref{spectrum}}
  \label{variability}
\end{figure}

\begin{figure*}
\includegraphics[width=155mm]{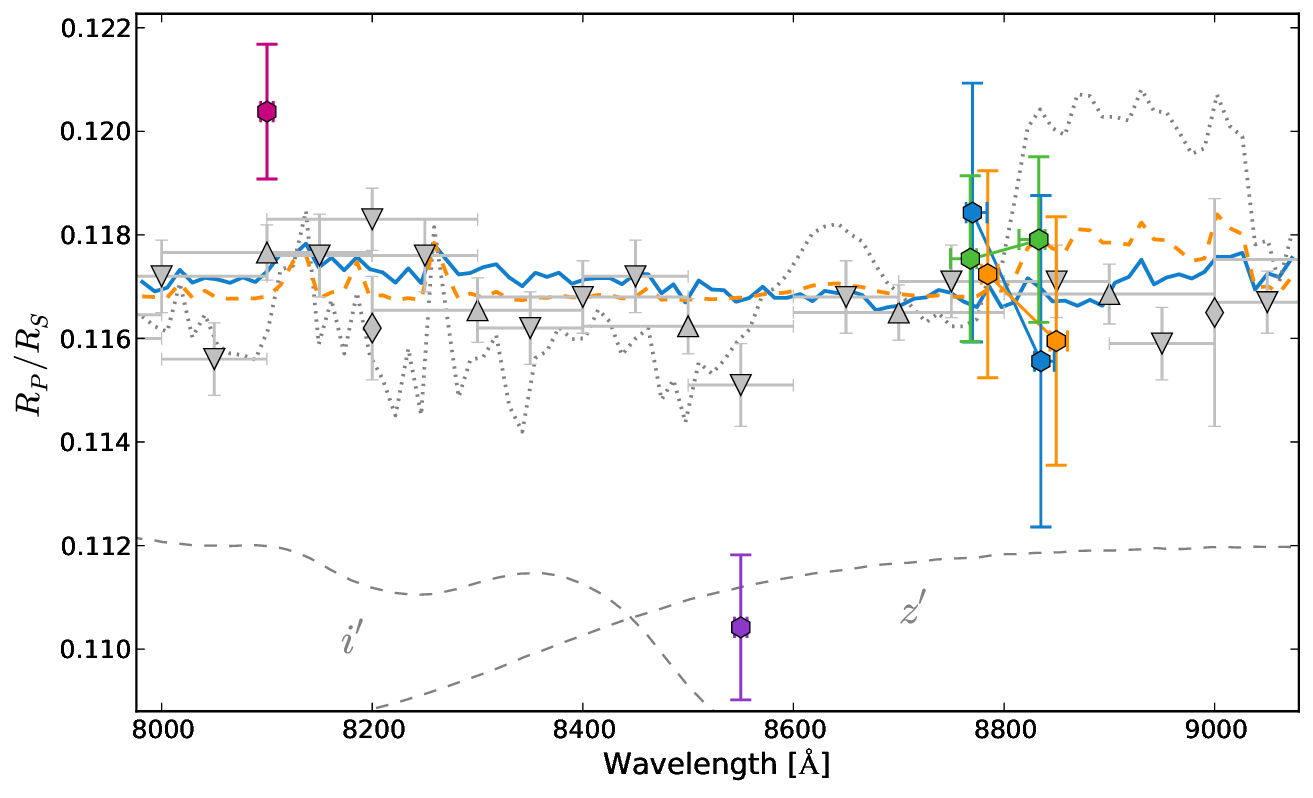}
 \caption{The combined transmission spectrum of GJ 1214b with data from \protect\cite{bean10} (grey upward triangles) and \protect\cite{bean11} (grey downward triangles) using the FORS instrument on the VLT (200~\AA\ bandpass), \protect\cite{demooij12} using the I-filter with the WFC on the Isaac Newton Telescope and z filter with the GROND instrument on the 2.2 m MPI/ESO telescope (grey diamong) and this study (multi-coloured hexagonal markers, 12~\AA\ bandpass). The horizontal error bars represent the width of the photometric band. The transmission spectra are from \protect\cite{morley13} and have been binned into 12~\AA\ bins for clarity. The observations are shown alongside a 100\% water atmosphere model (solid blue line), a best-fitting cloud-free 50$\times$ solar composition atmosphere model with an efficient heat distribution (grey dotted line) and a worst-fit cloudy (KCl and ZnS) 50$\times$ solar composition atmosphere model with a low sedimentation efficiency of $f_{\rm{sed}}=0.1$ and a efficient heat distribution (orange dashed line). A close-up of the methane probing region is shown in Fig.~\ref{models_crop}.}
  \label{spectrum}
\end{figure*}

The resulting light curves with their corresponding best-fit models for each transit observation are shown in Fig. \ref{transits-sing} and Fig. \ref{transits-colon}. The measured radius ratios are compared to atmospheric models by \cite{morley13} and are shown in Fig.~\ref{spectrum} and \ref{models_crop}.

\subsection{Variability due to Stellar Activity}
\label{stellar-activity}

GJ 1214, an active M4.5 type star, has been shown to exhibit a $\sim2$\% peak-to-peak stellar flux variability in the wavelength range 715--1000~nm and a long rotation period on the order of 53 days \citep{berta11} based on three years of data from MEarth \citep{nutzman08}. This is equivalent to a difference in radius ratio of $\Delta(R_p/R_s)\sim0.001$ (using Eq. 7 from \citealt{desert112}). Compared to the equilibrium cloud model atmosphere which includes methane, detailed in \S~\ref{models}, we would expect an increase in the radius ratio of the broad methane absorption band at 8800~--~9000\AA\ to be $R_p/R_s \sim0.002$ at the resolution of our measurements. As such it is necessary to consider the impact of stellar variability and star spots.

The stellar activity can affect the transit depth by means of unocculted star spots, which increase the transit depth, or by the presence of occulted star spots, which lead to an underestimation of the transit depth. As the star spot coverage changes due to the evolution of the spots and stellar rotation, it is possible that small differences in the transit depths are measured at different epochs. To limit the effects of stellar activity it is essential that the different wavelength observations are done at, or close to, the same time. In this study, the light curves that probed the methane feature were acquired over three transits by alternating the tuneable filters between two wavelengths. This gave a total of 6 individual transits around the methane feature (see Fig.~\ref{transits-colon}), two at $8770$ \AA\ and one at 8784.5 \AA\ (off-methane) and two at $8835$ \AA\ and 8849.6 \AA\ (on-methane).

The transit light curve at 8550 \AA\ taken on the 2$^{\rm{nd}}$ of June 2010 shows a considerably shallower transit depth inconsistent at the $\sim$3~$\sigma$ significance level with previously measured radius ratios by \citep{bean11} . Although the cause of this is unknown, it is possible the shallower light curve is partly the result of a below average number of star spots on the surface of GJ 1214, causing a shallower transit depth to be observed. No evidence for the presence of an occulted star spot is present in the data. \cite{carter11} also observed a shallower transit of GJ~1214b 3 nights later, during the morning of the $6^{th}$ of June 2010 (see Fig.~\ref{variability} near BJD-2455260 + 90). Considering that the estimated rotation speed of GJ 1214b is of the order of 53 days, it is likely that both observations were affected by a lower number of star spots than usual.

\subsection{The impact of the observed wavelength drifts}
\label{wave_shift}

\begin{figure}
\includegraphics[width=84mm]{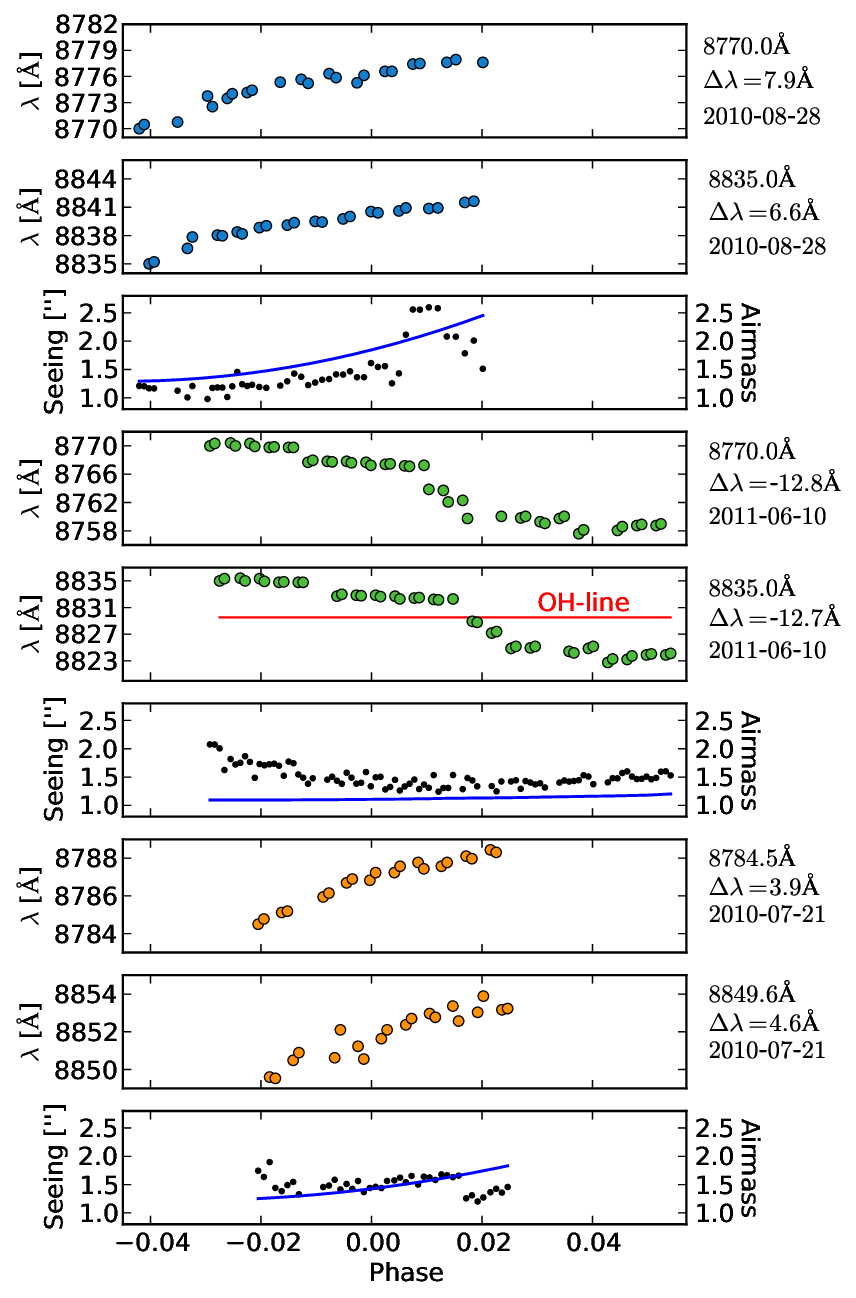}
 \caption{The panels above show the drift in wavelengths during each observation together with the changes in seeing and airmass. The seeing is indicated by black points (seeing scale on the left) whilst airmass is represented by a blue solid line in the same panel (airmass scale on the right). The red line shown in the fifth panel from the top shows the location of the OH sky line doublet near 8829.5~\AA. The drift in wavelength was calculated using Eq.~\ref{ring-eq}.}
  \label{shift}
\end{figure}

The change in sky line position during the course of the observations is indicative of a change in wavelength. By measuring the position of the sky lines in the images the corresponding drift in wavelength was estimated following Eq.~\ref{ring-eq} with the drifts shown in Fig.~\ref{shift}. The most significant wavelength drift is seen in the 10$^{\rm{th}}$ of June 2011 observations where a wavelength change of $\sim13$~\AA\ was observed. No detectable sky lines were observed during the 17$^{\rm{th}}$ of August 2010 and 2$^{\rm{nd}}$ of June 2010 observations. 

Although every attempt was made to tune the filter to a wavelength absent of strong sky lines it is likely that the sky lines still affect the data despite the sky background having been subtracted. In particular the OH-emission line doublet at 8829.514~\AA\ and 8829.525~\AA\ \citep{rousselot00} has likely interacted with the 8835~\AA\ observations conducted on the 10$^{\rm{th}}$ of June 2011. The 8770\AA\ observations, done the same night are not affected by a similar shift in wavelength which suggests the OH-emission line is likely causing the systematic relative flux variations seen in the 8835~\AA\ observations (Fig.~\ref{transits-colon-raw}). Shown in Fig.~\ref{shift} (fifth panel from the top) the observed wavelength drifted towards shorter wavelengths during the night, with the OH line (red line) causing an increase in the relative flux. This is clearly seen in the raw transit light curve shown in Fig.~\ref{transits-colon-raw} (middle green light curve on the right).

\begin{figure}
\includegraphics[width=84mm]{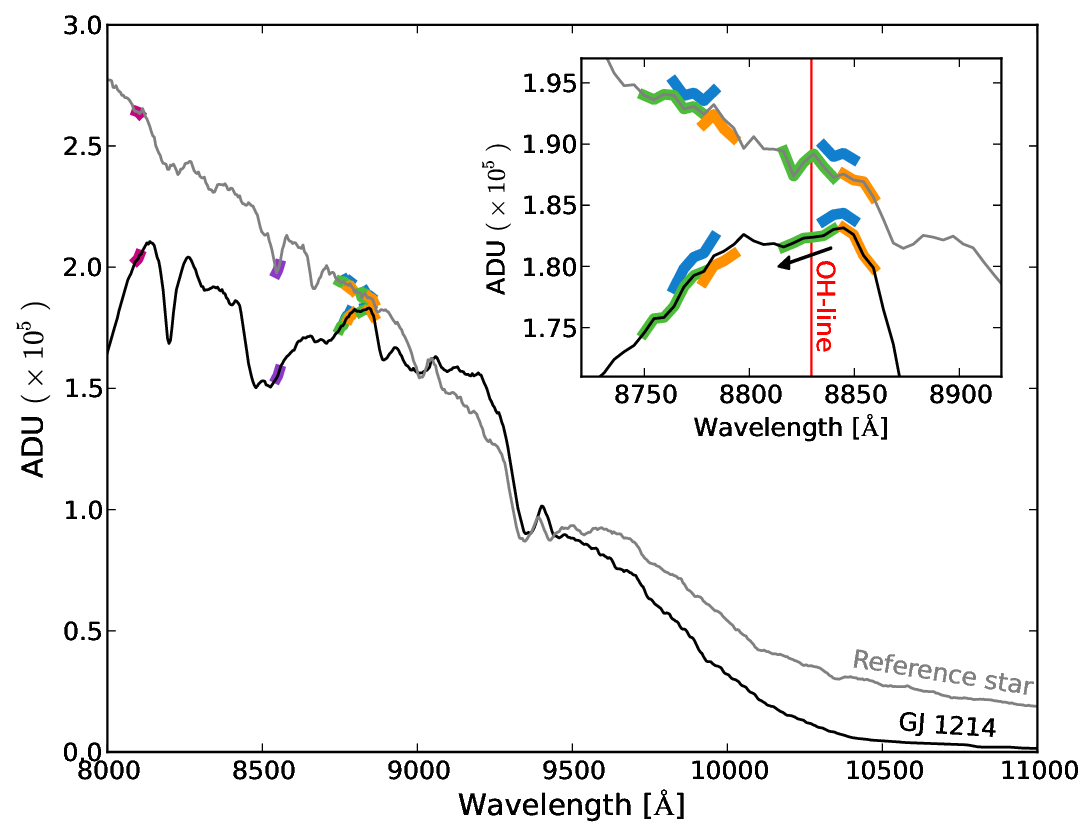}
 \caption{The spectrum of GJ1214 (black) and one of the reference stars with coordinates $17^h 15^m 19.42^s$ $04\degree 56\arcmin 32.6\arcsec$ (grey). The colours indicate the wavelength probed by each filer (including drifts). The reference spectrum has been multiplied by a factor 9.2 to allow for a comparison in the sub-window. The OH doublet at the vacuum measured wavelengths 8829.514~\AA\ and 8829.525~\AA\ is shown in red. Only the 8835~\AA\ observations on the 10$^{\rm{th}}$ of June 2011 cross this line completely with the black arrow indicating the direction of the wavelength drift.}
  \label{stellar_spec}
\end{figure}

With the sky-rings subtracted before performing aperture photometry only very small sky-ring residuals are left in the images. It was of interest to investigate if other systematics were introduced by the wavelength drift sampling different parts of the spectrum of GJ 1214 or the spectra of one of the reference stars. Having previously conducted spectroscopic observations of GJ 1214 with the GTC telescope using the R500R grism and a 10\arcsec~slit on the 25$^{\rm{th}}$ of July 2012, we compared the wavelength drifts with the spectra of GJ1214 and one of the reference stars used. The two spectra which have a resolution of about $R\sim250$ are shown in Fig.~\ref{stellar_spec} with a rescaled view of the methane-probing region shown in the sub-window located towards the upper right of the plot. Due to the nature of the long slit, only one other reference star could be fit on the slit. Since no major absorption lines were crossed it is unlikely that the systematic wavelength trends are caused by different parts of the spectrum of GJ 1214 or the reference star being sampled.

\subsection{Probing the methane feature}
\label{methane}

\begin{figure}
\includegraphics[width=84mm]{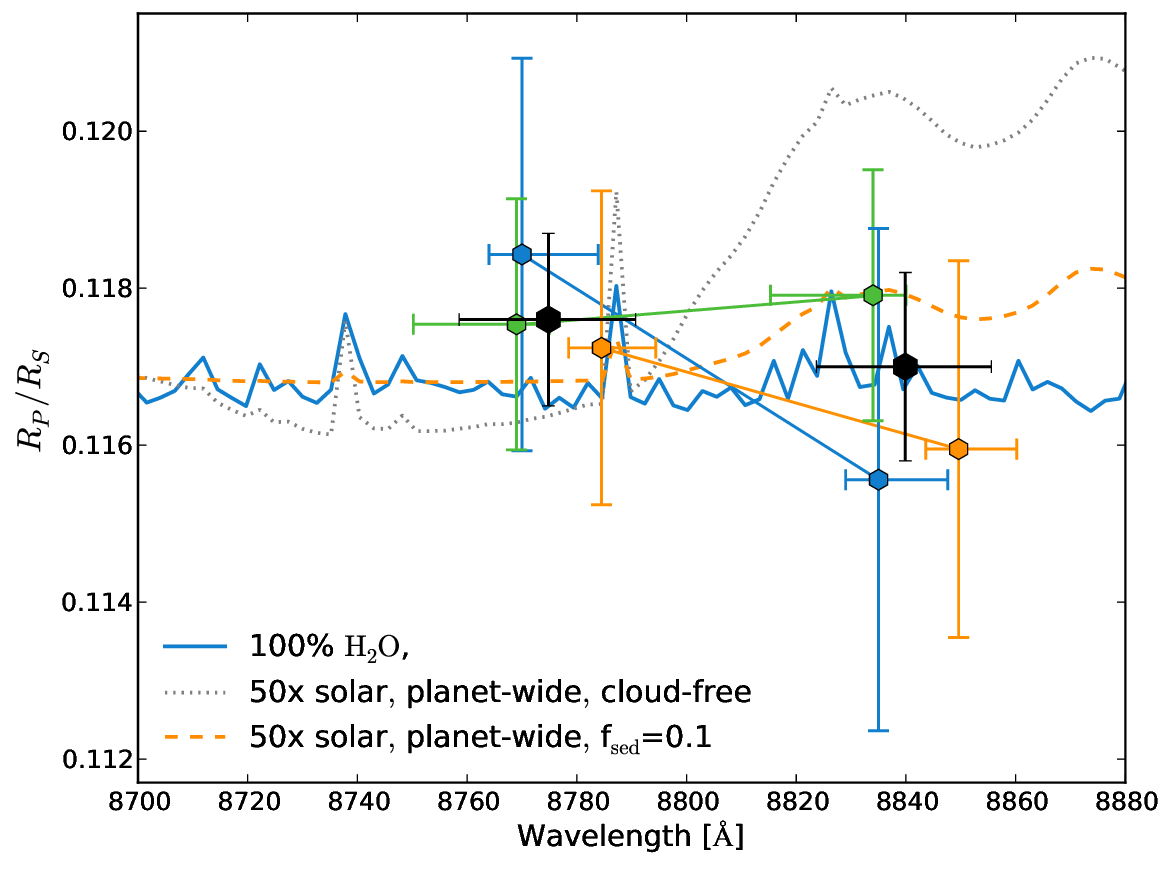}
 \caption{The weighted average of the on-methane and off-methane radius ratios (black points) compared to unbinned model spectra of a 100\% water atmosphere model (blue solid line), a best-fit cloud-free 50$\times$ solar composition atmosphere model with an efficient heat distribution (grey dotted line) and a cloudy (KCl and ZnS) 50$\times$ solar composition atmosphere model with a low sedimentation efficiency of $f_{\rm{sed}}=0.1$ and a efficient heat distribution (orange dashed line). The coloured points represents the radius ratios from the individual transits. For a larger overview of the transmission spectrum see Fig.~\ref{spectrum}.}
  \label{models_crop}
\end{figure}

Here we compare our five observed transits of GJ 1214b with theoretical models presented in \cite{morley13} to investigate the nature of GJ 1214b's atmosphere, in particular, to look for extra absorption due to methane (see Fig.~\ref{spectrum} and \ref{models_crop}). Using tuneable narrowband filters, we are able to probe the planets atmosphere at a higher spectral resolution (R~600--750) than would otherwise be possible using standard photometric filters. 

The possibility of a methane feature is explored by comparing the difference in radius ratios between the on-methane, off-methane observations each done on the same night. For the observations done on the 21$^{\rm{st}}$ of July 2010, the difference in radius ratios were found to be $\Delta R = -0.0013 \pm 0.0031$, for the 28$^{\rm{th}}$ of August 2010, $\Delta R = -0.0029 \pm 0.0041$, and for the 10$^{\rm{th}}$ of June 2011,  $\Delta R = 0.0004 \pm 0.0023$. The weighted average of the difference in radii from all three nights are calculated to be $\Delta \overline{R} = -0.0007 \pm 0.0017$ which in terms of scale heights ($H$) is expressed as $\Delta \overline{R} \simeq -0.5 \pm 1.2~H$, using $H/R_s = 0.0014$ and assuming a hydrogen dominated atmosphere. We therefore detect no increase across a possible methane feature. A close up of the probed methane region together with the weighted average of the on and off-methane planet-to-star radius ratios are show in Fig.~\ref{models_crop}.

\subsection{Atmosphere models}
\label{models}

We compare our observations across the methane feature to nine different atmosphere models, which include a 100\% water atmosphere model, four cloud-free models and four cloudy models from the model suites of \cite{morley13}. Each of the cloudy and cloud-free models consist of solar (1$\times$) and super solar metallicities (50$\times$) and a variety of T-P profiles. We find that our water and equilibrium cloud models (KCl and ZnS) fit the data well due to a large area of parameter space being allowed and are therefore not able to rule out any of the water or cloudy models. We show the worst fitting cloudy model, a 50$\times$ solar composition atmosphere with a low sedimentation efficiency of $f_{\rm{sed}}=0.1$ and a efficient heat distribution in Fig.~\ref{spectrum}. We are able to rule out all the cloud-free atmosphere models at the $>2.7\sigma$-level and show the best-fitting cloud-free models together with the worst-fitting cloudy-model in Fig.~\ref{spectrum}.

A muted methane feature would require the presence of optically thicker clouds or a large mean molecular weight atmosphere which would be hard to detect such as a water dominated atmosphere. For a cloudy H-rich model to obscure the transmission spectrum, the clouds have to be optically thick high in the atmosphere where transmission spectroscopy probes ($\sim$10$^{-3}$~bar). An inefficient heat redistribution would causes the P--T profile to be warmer, which would cause the profile to cross the condensation curve higher in the atmosphere. In addition, a low sedimentation efficiency would create a more vertically extended cloud. We note that photochemical models which include the formation of naturally forming photochemical hazes high in the atmosphere can also mute the transmission spectrum of GJ 1214b.

\cite{croll11} noted that the increased transit depth measured in the $K_{\rm{s}}$-band could be due to an opacity source such as methane or water, requiring hazes in the atmosphere. These measurements could also be explained by the observed spectral features being more complicated than current models predict. With a detection sensitivity on the order of a scale height at high spectral resolution, we are capable of excluding the presence of a methane absorption band spanning multiple scale heights, which higher resolution models might reveal in the future.

\section{Conclusions}
\label{conclusion}

We present GTC OSIRIS narrowband observations of five transits of GJ 1214b, three of which probe the presence of methane at two near simultaneously obtained wavelengths. We do not find no increase in radius ratios across the possible methane feature with a planet-to-star radius ratio, $\Delta \overline{R} = -0.0007 \pm 0.0017$, across the feature. This corresponds to an increase in scale height of $ -0.5 \pm 1.2~H$ assuming a hydrogen dominated atmosphere. We are therefore not able to rule out any of our water and cloud based models. Cloud-free models generally provide poor fits to the data. Even the best fitting cloud free model assuming a 50$\times$ solar composition atmosphere with an efficient heat distribution can be rejected at the 2.7$\sigma$ confidence level form our data alone. The results, which are compatible with previous results of a largely flat transmission spectrum, do not rule out the possibility of a high altitude haze or a water dominated atmosphere in the atmosphere of GJ 1214b, but do rule out methane features spanning multiple scale heights. Observations around the methane absorption band are predominantly limited by low cadence observations and sky emission lines in Earth's atmosphere affecting the photometric quality, making the determination of the systematic noise challenging. With tuneable filters capable of high resolution measurements (R $\approx$ 600-750) there is currently a need for high resolution methane models below 1 $\mu$m.

\section*{Acknowledgments}

We thank the entire GTC staff and in particular Antonio Cabrera Lavers and Robert C. Morehead for their help with conducting these observations. This work is based on observations made with the Gran Telescopio Canarias (GTC), installed in the Spanish Observatorio del Roque de los Muchachos of the Instituto de Astrof\'isica de Canarias on the island of La Palma. The GTC is a joint initiative of Spain (led by the Instituto de Astrof\'isica de Canarias), the University of Florida and Mexico, including the Instituto de Astronom\'ia de la Universidad Nacional Aut\'onoma de M\'exico (IA-UNAM) and Instituto Nacional de Astrof\'isica, Optica y Electr\'onica (INAOE). P.A.W, D.K.S and A.R.P acknowledges support from STFC. K.D.C. and R.C.M. were supported by NSF Graduate Research Fellowships. G.E.B. acknowledges support from STScI through grants HST- GO-12473.01-A. E.B.F acknowledges support from the Center for Exoplanets and Habitable Worlds is supported by the Pennsylvania State University, the Eberly College of Science, and the Pennsylvania Space Grant Consortium. F.P. is grateful for the STFC grant and Halliday fellowship (ST/F011083/1). This work was also aided by the National Geographic Society's Young Explorers Grant, awarded to K.D.C. The authors would like to acknowledge the anonymous referee for their useful comments.

\appendix

\section[]{Short baselines and the radius ratio uncertainty}
\label{appendix}
\begin{figure}
\includegraphics[width=84mm]{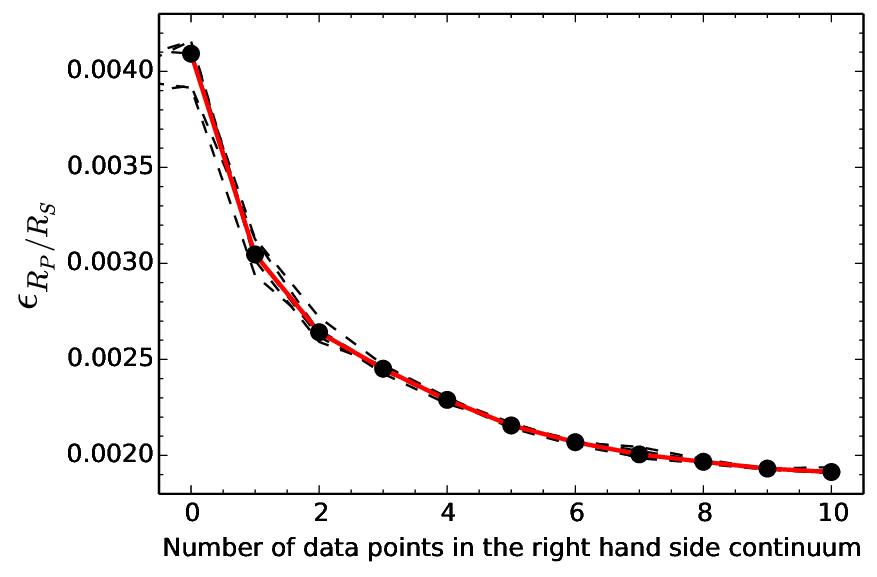}
 \caption{The uncertainties on the radius ratios as a function of the number of points in the right hand continuum. The results from the five light curves are shown as dashed lines with their median value represented as a red solid line. The data consists of the 2010 August 28 8770~\AA\ transit data with additional synthetic data points added to the right hand continuum. As the the number of data points in the right hand side continuum decrease, the uncertainty on the radius ratio given by the MCMC method increases following a power law.}
  \label{mcmc_uncertainties}
\end{figure}

With only a few data points present in the continuum of some of the transits, it was of interest to explore the effects of a short baseline on the uncertainty of the radius ratio as calculated using the MCMC method. To asses the relationship, we generated five light curves each consisting of the the 2010 August 28 8770  transit data with additional synthetic data points added to the right hand continuum. For each of the five light curves, a series of MCMC chains, each consisting of 500,000 steps (trimming away the first 50,000 points), were calculated iteratively removing one point from the right hand continuum before calculating a new chain. This process was repeated for the five light curves with the resulting radius ratio uncertainties subsequently median combined. As shown in Fig. \ref{mcmc_uncertainties}, as points are removed from the continuum the radius ratio uncertainties as given by the MCMC method increase following a power law. The relationship is further verified by our observations when comparing the derived uncertainties on the radius ratio between the 2010-08-28 and the 2011-06-10 transits, which were done at the same wavelengths and show a consistent decrease in uncertainties with the addition of more points in the post-egress continuum.

\bibliographystyle{mn2e}
\setlength{\bibhang}{2.0em}
\setlength\labelwidth{0.0em}
\bibliography{references}

\label{lastpage}

\end{document}